\documentclass[superscriptaddress]{revtex4-1}
\usepackage{graphicx}
\usepackage{color, soul}
\usepackage{amsmath,amsthm,amssymb,latexsym}

\begin{document}



\title{Quantum ferromagnet in the proximity of the tricritical point}

\author{Petr Opletal}
\author{Jan Prokle\v{s}ka}
\email{prokles@mag.mff.cuni.cz}
\author{Jaroslav Valenta}
\author{Petr Proschek}
\author{Vladim\'{\i}r Tk\'{a}\v{c}}
\author{R\'{o}bert Tarasenko}
\author{Marie B\v{e}hounkov\'{a}}
\affiliation{Charles University, Faculty of Mathematics and Physics, Ke Karlovu 3, Praha 2, 121 16, Czech Republic}

\author{\v{S}\'arka Matou\v{s}kov\'a}
\affiliation{Institute of Geology, Czech Academy of Science v.v.i., Rozvojova 269, Prague 6, 165 02, Czech Republic}

\author{Mohsen M. Abd-Elmeguid}
\affiliation{Charles University, Faculty of Mathematics and Physics, Ke Karlovu 3, Praha 2, 121 16, Czech Republic}
\affiliation{II. Physikalisches Institut, Universit\"{a}t zu K\"{o}ln, 50937 K\"{o}ln, Germany}

\author{Vladim\'{\i}r Sechovsk\'{y}}
\affiliation{Charles University, Faculty of Mathematics and Physics, Ke Karlovu 3, Praha 2, 121 16, Czech Republic}


\begin{abstract}

Echoes of quantum phase transitions (QPTs) at finite temperatures are theoretically and experimentally challenging and unexplored topics. Particularly in metallic quantum ferromagnets the experimental investigations are hampered 
by an intricate preparation of sufficiently pure samples and the access to the proper coordinates in parameter space. The present study shows that it is possible to tune a specific system at easily accessible conditions to the 
vicinity of its quantum phase transition. The physics is demonstrated on Ru-doped UCoAl, driven by pressure or substitution to and across the tricritical point and follows the first-order transition line to the theoretically 
presumed QPT. These findings open the possibilities for further in-depth studies of classical and quantum critical phenomena at easily reachable conditions.

\end{abstract}

\maketitle


Quantum Phase Transitions (QPTs) have become an important research area of different fields of modern condensed matter physics such as metal-to-insulator \cite{Frandsen2016,Lahoud2014}, superconductor-to-insulator 
\cite{Pradhan2015}, superconductor-to-metal \cite{Schneider2012} and magnetic or electric order--disorder transitions~\cite{Frandsen2016,Brando2016,Lohneysen2007,Tokiwa2014,Rowley2014,Pfleiderer2001}.
QPTs occur at zero temperature, and can be 
triggered by varying a non-thermal control parameter as external pressure, chemical composition or magnetic field and driven by a corresponding change of quantum fluctuations between phases. Among these, magnetic quantum phase 
transitions are of special interest as they can have different physical origins and their nature depends on the type of the magnetically ordered state. While quantum critical points and quantum criticality in many itinerant and 
localized antiferromagnetic systems have been identified and well investigated, for ferromagnetic systems theory predicts a first-order QPTs for sufficiently clean systems rather than critical points \cite{Belitz2005,Belitz1999}, but 
available experimental evidence is rather limited.
For weak itinerant ferromagnets, it has been shown~\cite{Belitz2005,Belitz1999}~that, at sufficiently low temperatures, the phase transition is generically first order with 
a tricritical point (TCP), which separates the 
line of first-order transitions at low temperatures from the line of the second-order transitions at higher temperatures. Upon application of an external magnetic field tricritical wings appear, with theirs wing-tip 
points located at zero temperature (quantum-wing critical point, QWCP). The line of crossover temperatures $T_0$ between the TCP and the QWCP is characterised by a change of the order of the transition from first to second order.

Regarding the experimental investigation to explore the evolution of QPTs in metallic quantum ferromagnets and the theoretically predicted phase diagrams, the investigated systems are almost exclusively driven to criticality by 
hydrostatic pressure (see e.g. \cite{Brando2016}), with limited use of potential probing techniques, making difficult to distinguish between an intrinsically continuous second-order transition and an experimentally averaged and 
smeared out first-order one.

In the present work we have chosen UCoAl (hexagonal ZrNiAl-type structure with space group P-62m) whose ground state is paramagnetic at ambient pressure and zero magnetic field but located in the immediate proximity to the TCP 
and close to a ferromagnetic instability and thus is very sensitive to magnetic field, pressure, and alloying \cite{Sechovsky1998,Mushnikov1999,Andreev1997a,Aoki2011,Kimura2015}. UCoAl is therefore an excellent candidate for such 
studies compared to related U$TX$ compounds, where the QPT may be hidden in the superconducting dome (URhGe \cite{Aoki2001}, UCoGe \cite{Huy2007}) or accessible with extreme difficulties (URhAl). In the case of 
UCoAl, the QPT is expected to be positioned at small negative hydrostatic pressures and it is known that the ferromagnetic ground state can be achieved by doping with a small amount of transition metal. However, the 
substitutional tuning is known to introduce disorder, causing the suppression of TCP and thereby leads to the occurrence of a continuous transition to the lowest temperatures \cite{Brando2016}. In order to 
clarify the presence of the QPT and to investigate the vicinity of the TCP we prepared high quality single-crystaline samples of UCoAl with sub-percent doping by Ru of 0.5 and 1\% to avoid the effect of disorder.

Here we show that the UCo$_{1-x}$Ru$_x$Al system can be driven to and across the QPT by weak Ru doping or by hydrostatic pressure and we were able to resolve the details of the phase diagram in the 
vicinity of the TCP and the presence of a regime where long-range ferromagnetism and paramagnetism coexist due to the first-order transition below TCP. The results are fully consistent with theoretical predictions and offer the 
possibilities for further in-depth studies of quantum critical phenomena at easily achievable conditions.


\section{Results}

\begin{figure}
  \centering
  \includegraphics[width=\textwidth]{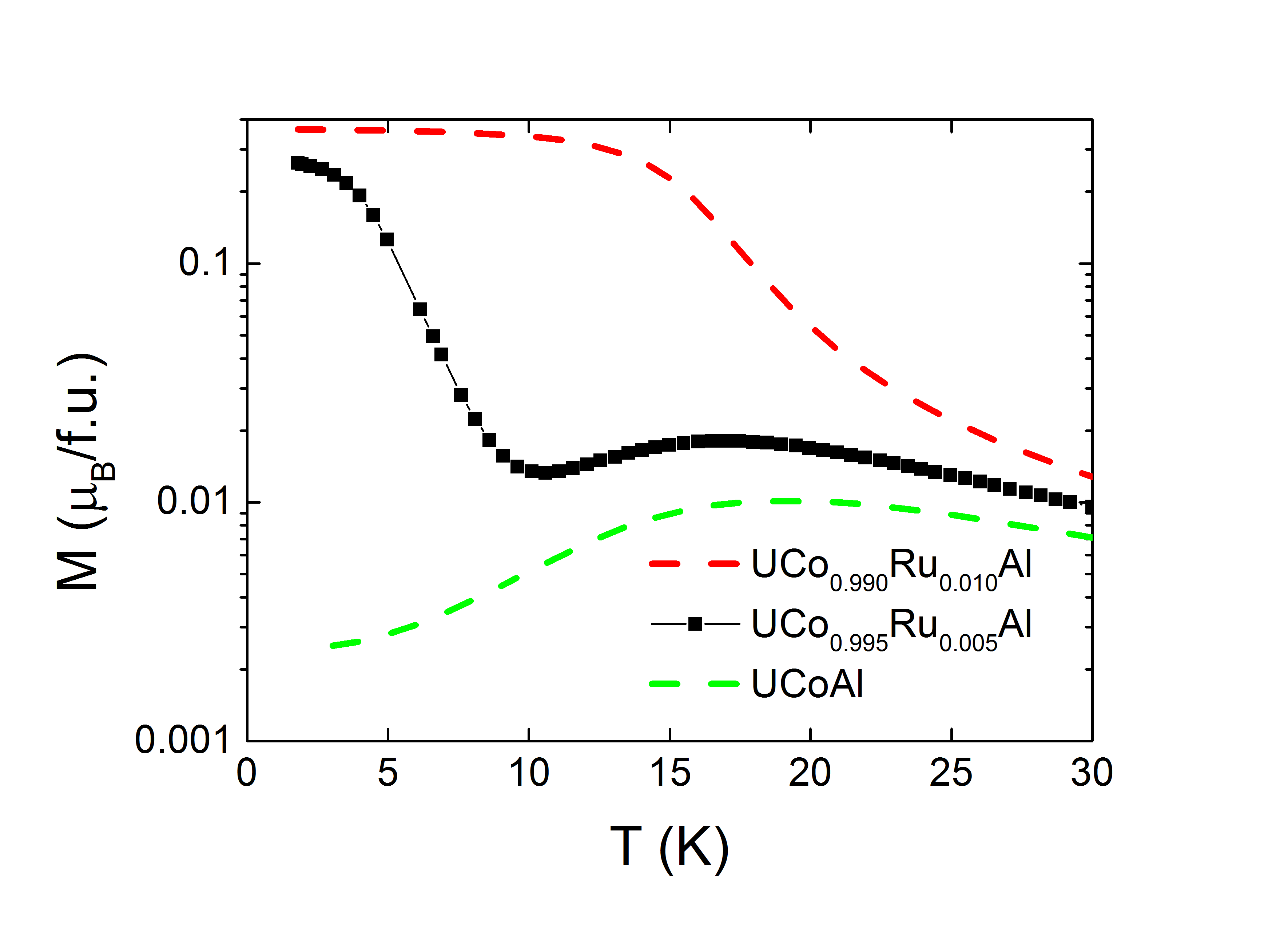}
  \caption{
Temperature dependence of the low-field (0.1~T applied along the c-axis) magnetization for UCo$_{0.995}$Ru$_{0.005}$Al, UCo$_{0.990}$Ru$_{0.010}$Al and UCoAl. Logarithmic scale (vertical axis) is used to visualize features 
connected with spin-fluctuations.} 
\label{F1}
\end{figure}

As we mentioned above, the undoped compound UCoAl is paramagnetic at ambient pressure and zero magnetic field but located in the immediate proximity to the TCP and close to a to ferromagnetic instability. It exhibits a strong 
magnetic anisotropy~\cite{Sechovsky1998} due to the overlap of 5f wave functions of neighbouring U atoms and hybridisation of the U 5f states with the valence states of $T$ and $X$. Such an overlap together with the 5f-electron 
orbital moment results in an Ising-like character of the magnetism with the easy-magnetization direction along the c-axis.

\begin{figure}
  \centering
  \includegraphics[width=\textwidth]{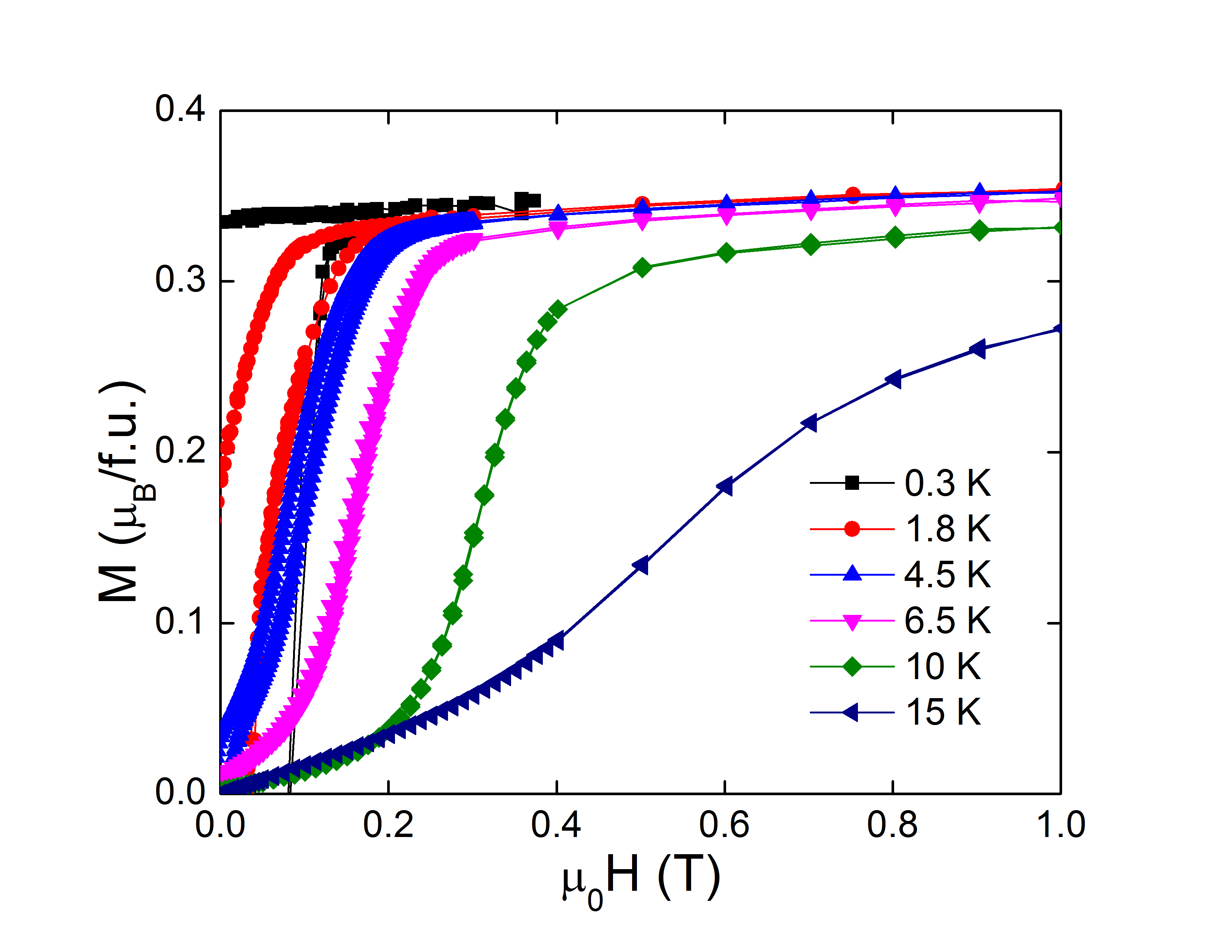}
  \caption{
Magnetization of a UCo$_{0.995}$Ru$_{0.005}$Al single crystal at selected temperatures measured  in a magnetic field applied along the c-axis.}
\label{F2}
\end{figure}

The effect of Ru doping on the magnetic behaviour of UCoAl is best seen in Fig.\ref{F1} which shows the temperature dependence of the low-field magnetization along the c-axis of the doped single crystal samples with 0.5\% and 1\% of Ru 
as compared to the UCoAl. Fig.~\ref{F2} displays the temperature dependence of the magnetization of UCo$_{0.995}$Ru$_{0.005}$Al single crystal at selected temperatures measured in a magnetic field applied along the c-axis.

As it is shown in Fig.\ref{F1} and Fig.~\ref{F2}, we observe in the sample doped with 0.5\% of Ru a shift of the metamagnetic transition at the respective temperatures to lower fields in comparison to pure UCoAl ($H_{\rm c}\approx 
0.7$~T) whereas the spin-fluctuation maximum~\cite{Yamada1993} in the low-field $M$ vs $T$ dependence is preserved (see Fig.~\ref{F1}). This indicates the presence of strong spin fluctuations associated with the metamagnetic 
behaviour similar to that in pure UCoAl.

In contrast to the behaviour of UCo$_{0.995}$Ru$_{0.005}$Al, UCo$_{0.990}$Ru$_{0.010}$Al behaves as a ferromagnet with a Curie temperature of 16~K, as determined from the temperature dependence of the low-field magnetization (see 
Fig.~\ref{F1}).  Here we observe neither a sign of a local maximum in the temperature dependence of the magnetic susceptibility nor metamagnetic behaviour.  Thus the two doped samples probe different magnetic regions of the phase 
diagram close to the TCP and thereby allows one to explore the change of nature of the magnetic state by tuning the system towards and across the QPT.

\begin{figure}
  \centering
  \includegraphics[width=\textwidth]{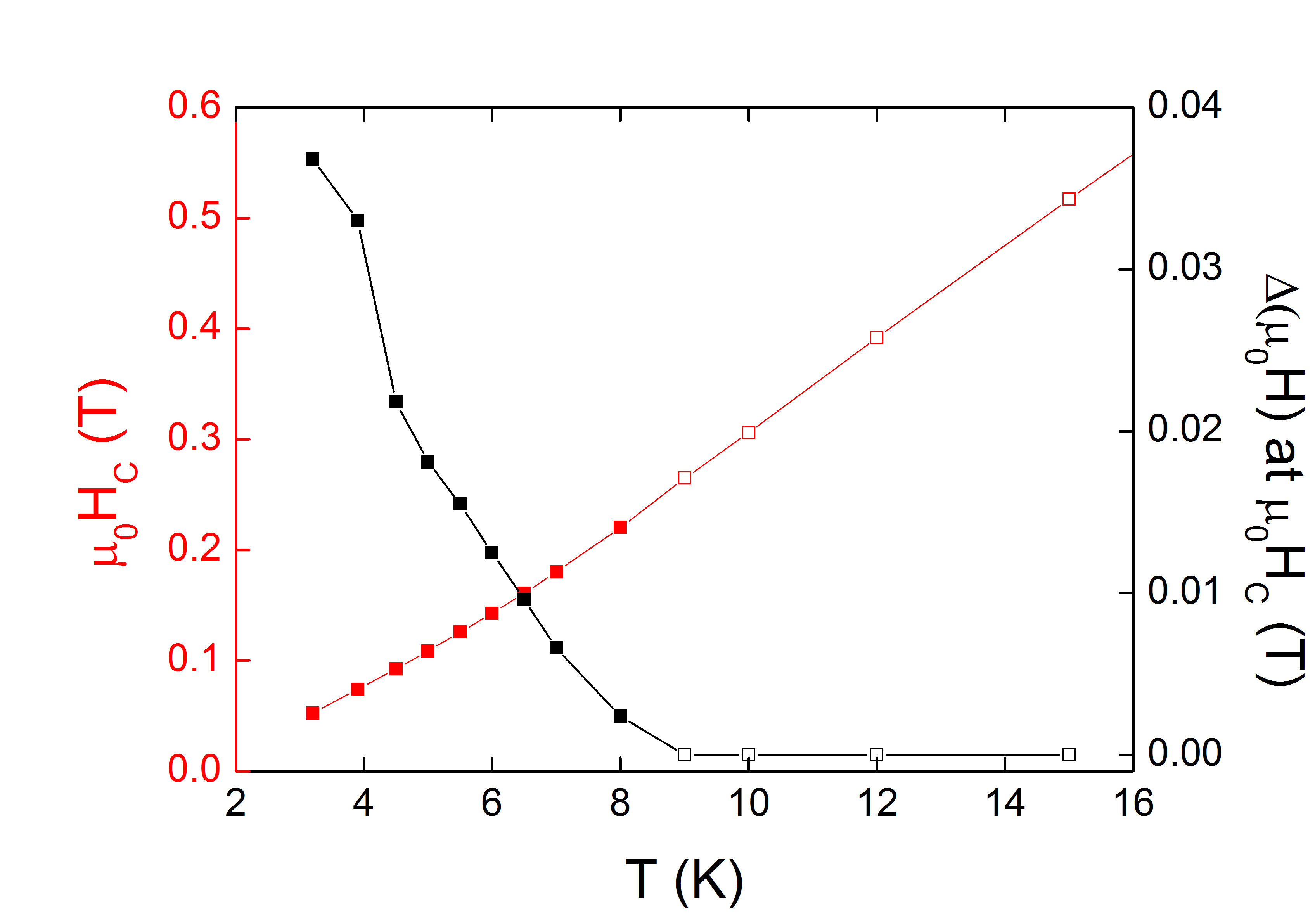}
  \caption{
Temperature dependence of critical magnetic field $\mu_0 H_{\rm c}$ and size of hysteresis $\Delta\mu_0 H_{\rm c}$ at $H_{\rm c}$ for UCo$_{0.995}$Ru$_{0.005}$Al for field applied along 
the c-axis. Filled symbols indicate fields at which a first-order transition is observed ($H_c$ determined as an average of field up and field down values), open symbols corresponds to a second-order transition.}
\label{F3}
\end{figure}

We first discuss the magnetic state of UCo$_{0.995}$Ru$_{0.005}$Al sample. The temperature dependence of critical magnetic field $\mu_0 H_{\rm c}$ and size of hysteresis $\Delta\mu_0 H_{\rm c}$ at $H_{\rm c}$ for field applied 
along the c-axis is shown in Fig.~\ref{F3}.
The hysteresis at the 
metamagnetic transition vanishes in the 9~K magnetization loop, allowing the determination of the crossover point at 0.3~T to be 9~K.
However, at low temperatures, a finite remnant magnetization (Fig.~\ref{F2}) emerges indicating ferromagnetic ordering. The remnant magnetization decays exponentially with temperature 
showing a tendency to saturation at low temperatures. Indeed, measurements at low temperatures ($T<2$~K) unveil the fully developed ferromagnetic loop below 0.6~K (Fig.~\ref{F2}). At 1.8~K, the coercive field of the
ferromagnetic loop and the critical field of the metamagnetic transition are comparable and the two phenomena cannot be distinguished. The ZFC and FC curves at 5~mT indicate ferromagnetic order in the vicinity of 4~K. In
 this temperature region, the critical magnetic field $H_{\rm c}$ is sufficiently larger than the applied magnetic field (see~Fig.~\ref{F3}) which allows us to exclude the influence of the metamagnetic transition.

\begin{figure}
  \centering
  \includegraphics[width=\textwidth]{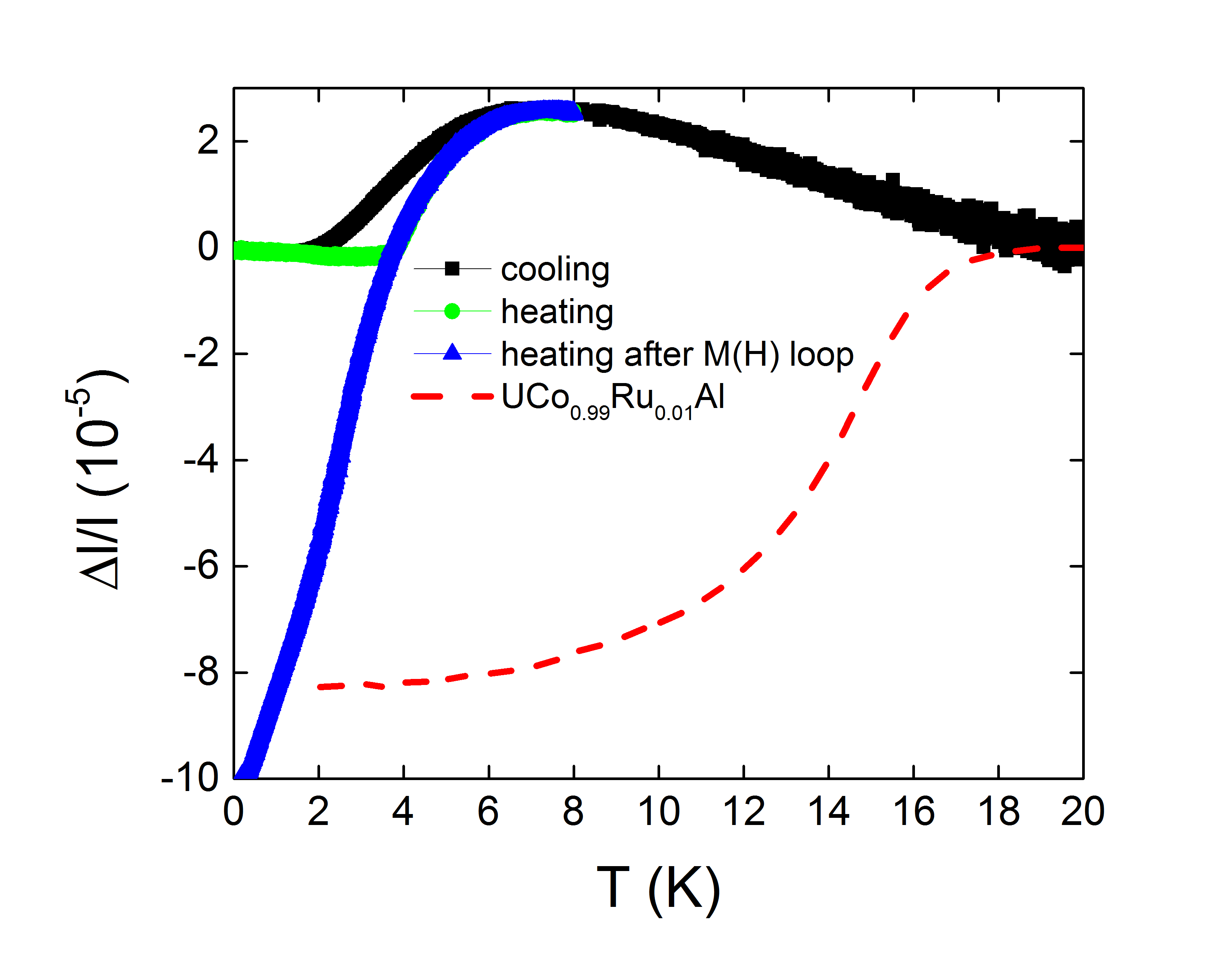}
  \caption{ Temperature dependence of thermal expansion of UCo$_{0.995}$Ru$_{0.005}$Al (symbols) and UCo$_{0.990}$Ru$_{0.010}$Al (dashed line). All measurement are done along the c-axis at zero field, measurement denoted 
by blue triangles was done after the crystal was fully magnetised at 100mK.}
\label{F4}
\end{figure}

In order to clarify the low-temperature behaviour of UCo$_{0.995}$Ru$_{0.005}$Al and the nature of its ground state, the temperature dependence of the thermal expansion along the c-axis was measured in zero magnetic field 
(Fig.~\ref{F4}). After initial cooling, we observe with increasing temperature a pronounced upturn at 3.8~K followed by a weak maximum in the vicinity of 7~K. With further heating, the sample contracts, reaching at 20~K 
approximately the same length as in the low temperature ordered state. There are remarkable connections between the observed features in the magnetization and in the thermal expansion data. The determined values of the ordering 
temperature agree very well, the maximum in the thermal expansion coincides with the disappearance of the ferromagnetic loop, whereas 20~K is the temperature where the magnetic susceptibility maximum is observed. Upon cooling 
down, the transition becomes smeared out and a wide hysteresis appears, starting at the thermal expansion maximum at 7~K down to 2~K, deeply in the ordered state. In order to unambiguously relate the transition at 3.6~K to 
ferromagnetism, the sample was fully magnetised at low temperatures (at 100~mK, a magnetic field of 1~T along the c-axis was applied) and the measurement during heating was repeated after field removal. Due to the remnant 
magnetostriction, the curve starts at negative values and approaches zero at 4~K, as a result of the domain-structure decomposition connected with the loss of long-range ordering. This allows us to conclude, that the compound exhibits 
a ferromagnetic ground state, yet metamagnetism (a feature intimately connected to the paramagnetic ground state of UCoAl) exists close to and above the Curie temperature. This indicates the presence of or close proximity of 
the sample to the tricriticality in the generalised phase diagram.

\begin{figure}
  \centering
  \includegraphics[width=\textwidth]{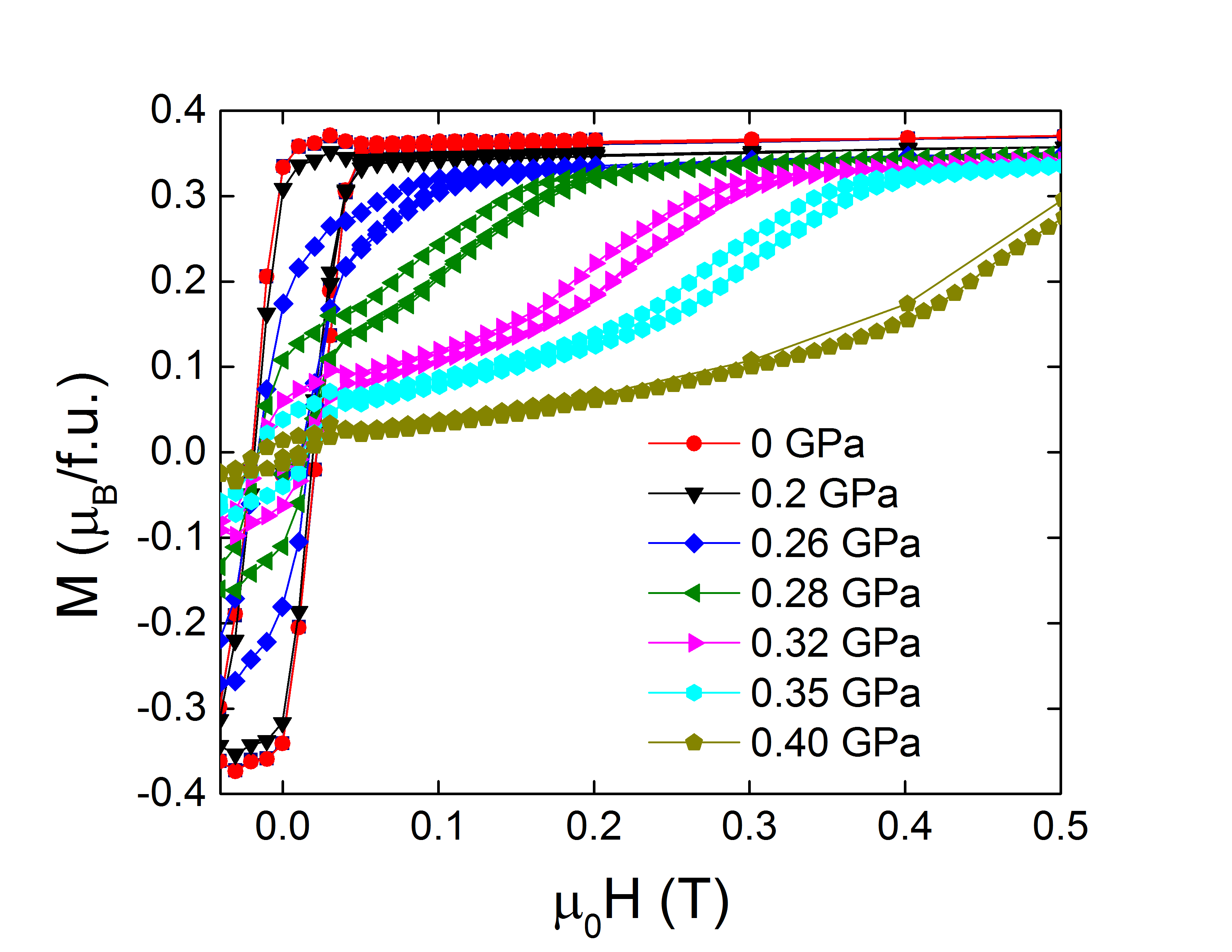}
  \caption{Magnetization at 5~K of a UCo$_{0.990}$Ru$_{0.010}$Al single crystal at selected hydrostatic pressures in a magnetic field applied along the c-axis. Obtained data were cut in order to properly visualize both coercivity 
and metamagnetism at once.}
\label{F5}
\end{figure}

\begin{figure}
  \centering
  \includegraphics[width=\textwidth]{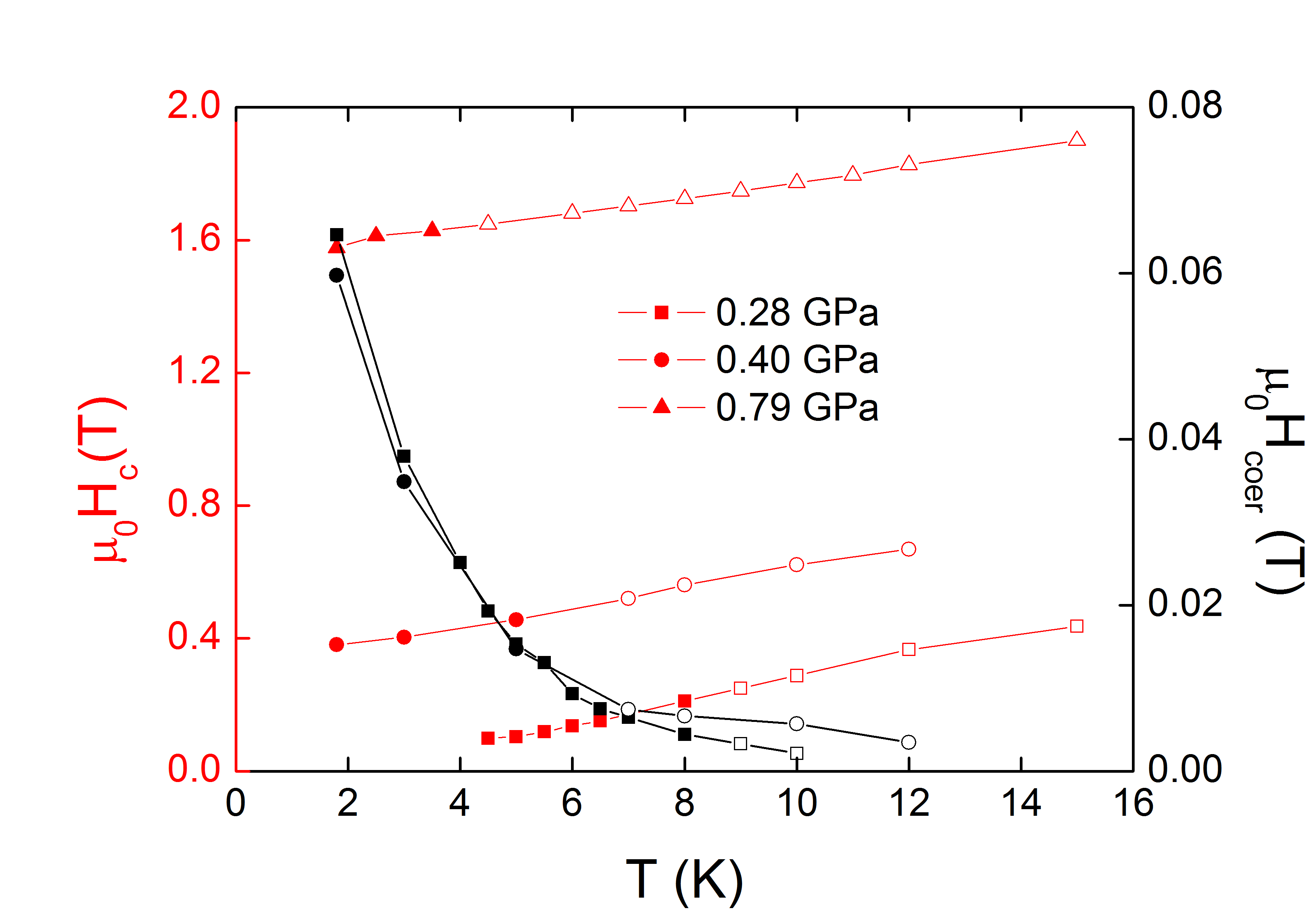}
  \caption{Temperature dependence of critical magnetic field $\mu_0 H_c$ of the metamagnetic transition and coercive field $\mu_0H_{\rm coer}$ of ferromagnetic portion of the curve for selected pressures for the 
UCo$_{0.990}$Ru$_{0.010}$Al sample for field applied along the c-axis, full symbols indicate the fields where first order transition is observed, open symbols indicate the presence of second order transition.}
\label{F6}
\end{figure}

\begin{figure}
  \centering
  \includegraphics[width=\textwidth]{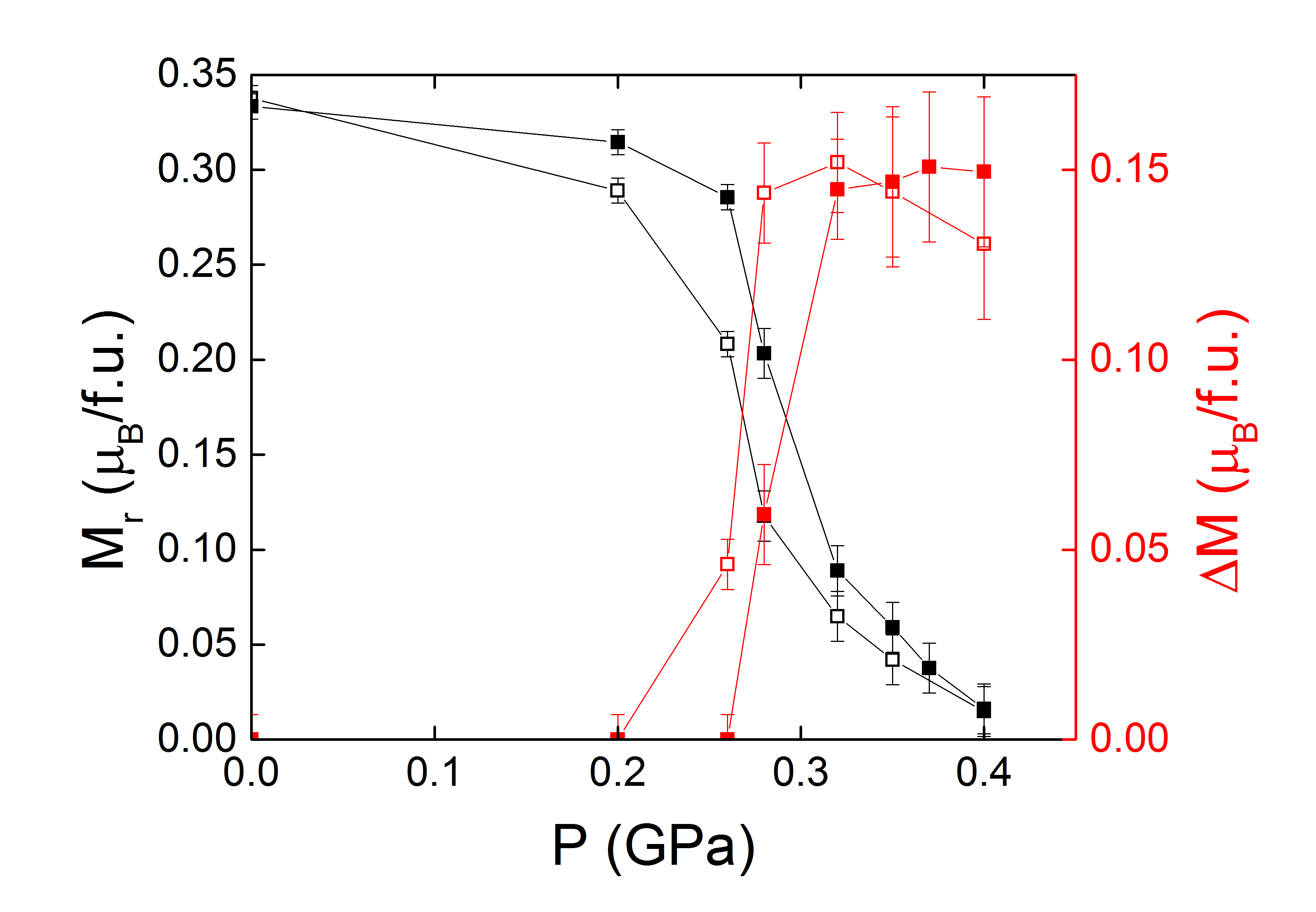}
  \caption{
Pressure dependence of the magnetization step $\Delta M$ at the metamagnetic transition and remnant magnetization $M_{\rm r}$ at two selected temperatures for the UCo$_{0.990}$Ru$_{0.010}$Al sample for field applied along the c-axis.}
\label{F7}
\end{figure}

To elucidate our findings that indicate the coexistence of 
 ferromagnetism and paramagnetism in the limited range of parameter space of 0.5~\% Ru-substitution, we have investigated the evolution of the ferromagnetic state of the sample doped with 1\% Ru under pressure. In other words 
tuning the ferromagnetic sample towards and across the expected QPT. 
The pressure dependence of the magnetization curves at 5~K of a UCo$_{0.990}$Ru$_{0.010}$Al single crystal at selected hydrostatic pressures in a magnetic field applied along the c-axis is shown in Fig.~\ref{F5}.
At low pressures (up to 0.2~GPa), the ground state remains ferromagnetic (see Fig.~\ref{F5} as an exemplary excerpt of obtained data), yet the Curie 
temperature is quickly suppressed with a rate of 25~K/GPa. With further increase of pressure, the change (decrease) of the Curie temperature levels off (cf. Fig.~\ref{F6}) at 0.26~GPa and 7~K. Simultaneously, one 
observes an additional curvature in the low-temperature magnetization (close to the coercive field, see Fig.~\ref{F5} at $p=0.26$~GPa) which becomes more pronounced at higher pressure or temperature. Schematically, one can 
capture the evolution of magnetic state with pressure by plotting the basic characteristics of the metamagnetism as has been done in previous Fig.~\ref{F3} and by adding the coercive field of the ferromagnetic part of the magnetization loop. The resulting 
Fig.~\ref{F6} clearly shows the main features of the magnetic states developing with pressure. It can be seen, that the temperature evolution of the coercive field (ferromagnetic quantifier) remains intact at these modest 
pressures, in agreement with the negligible pressure dependence of the Curie temperature, whereas the critical field of the metamagnetic transition (paramagnetic-ground-state quantifier) monotonously evolves. An estimation 
of the evolution of the critical field with respect to pressure gives 2.5~T/GPa, being comparable to the value reported for pure UCoAl~\cite{Andreev1997a}. This can be described as the presence of phase separation within a limited window of pressures, 
composed of an itinerant ferromagnet phase and a metamagnetic phase with identical characteristic and pressure-independent temperatures $T_{\rm C}=T_0=T_{\rm tr}\approx 7$~K. With further increasing pressure, ferromagnetism 
is quickly suppressed and a purely paramagnetic ground state exists. The behaviour is illustrated in Fig.~\ref{F7} in which one can clearly see the complementarity in the evolution of the remnant magnetization (related to the 
ferromagnetism) and the magnetization step at the metamagnetic transition (related to the paramagnetic part of the ground state). Based on the presented data, we locate the TCP for UCo$_{0.990}$Ru$_{0.010}$Al at 0.2~GPa and 
8~K.

In order to investigate the critical behaviour across the above described loss of long range ferromagnetism we have investigated the electrical resistivity of the UCo$_{0.990}$Ru$_{0.010}$Al sample under hydrostatic pressures as 
well. The obtained temperatures dependencies were evaluated assuming the $\rho=\rho_0+AT^n$ behaviour. At low pressures and low temperatures (in ferromagnetic ordered state) the resistivity follow a $T^2$ law as expected from the 
Fermi-liquid theory. With increasing pressure the characteristics abruptly changes – most notably the power drops to 3/2 and remains unchanged up to 0.45GPa (maximum applied pressure in the study). It should to be noted that the 
$\rho\sim T^{3/2}$ law is valid in extended temperature range up to 10K. Although the values of $A$ and $\rho_0$ are comparable to those of pure UCoAl~{\cite{Aoki2001}}, this is not true for the power law, where 5/3 power is observed in 
investigated isostructural U$TX$ compounds driven across the criticality~\mbox{\cite{Shimizu2015, Misek2017}}, including stoichiometric UCoAl itself. Noteworthy exceptions from this are slightly off-stoichiometric U(Co,Al) 
compounds~\mbox{\cite{Kolomiets1998}} or small Ru doping (UCo$_{0.9975}$Ru$_{0.0025}$Al, this study) having paramagnetic ground state and following $\rho\sim AT^{3/2}$ law at low temperatures.

\begin{figure}
  \centering
  \includegraphics[width=\textwidth]{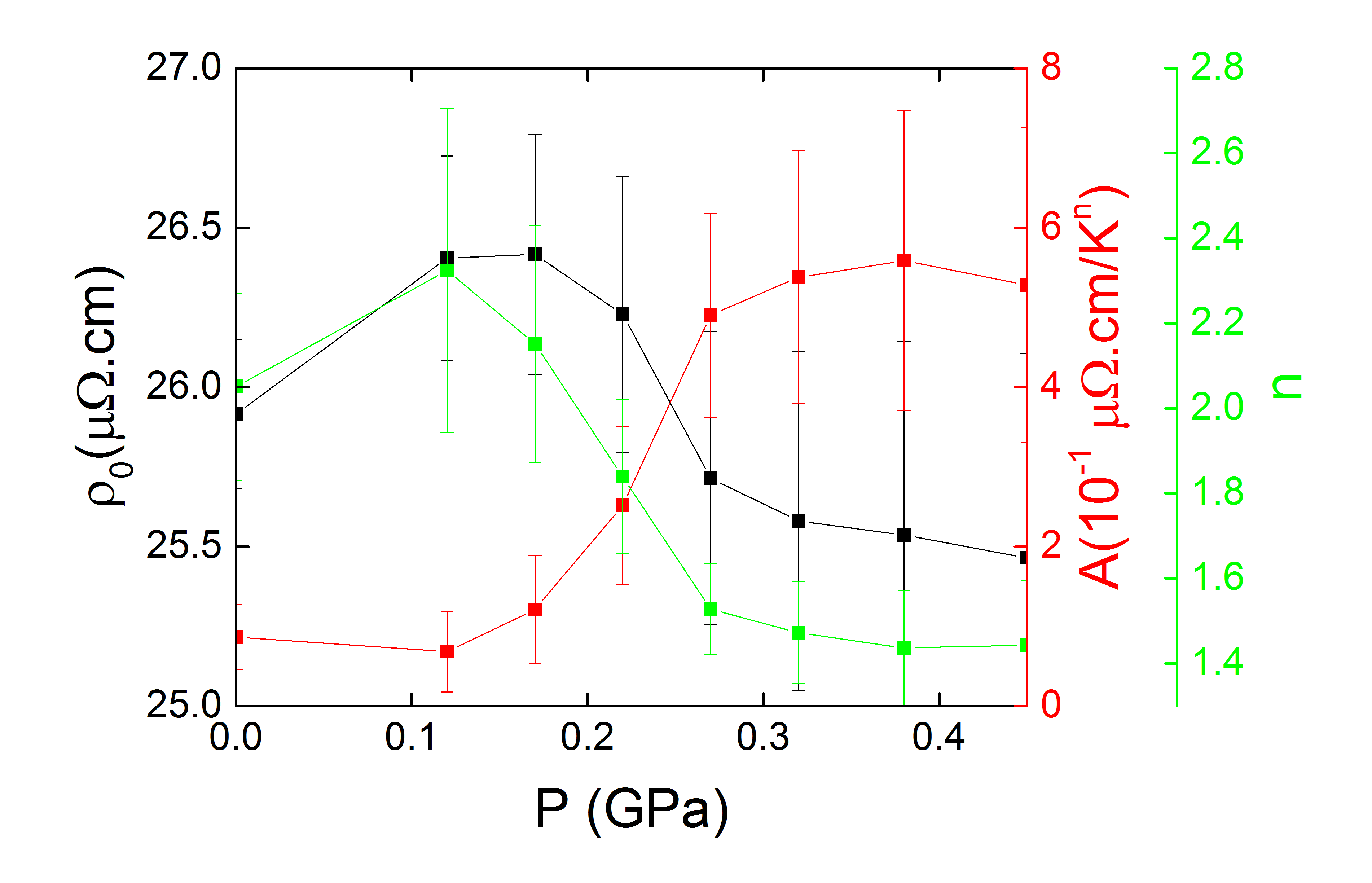}
  \caption{
The pressure dependence of residual resistivity ($\rho_0$), $A$ coefficient and power $n$ as evaluated from the temperature dependencies of the electrical resistivity in the case of the  UCo$_{0.990}$Ru$_{0.010}$Al sample.}
\label{F8}
\end{figure}

\section{Discussion}

\begin{figure}
  \centering
  \includegraphics[width=\textwidth]{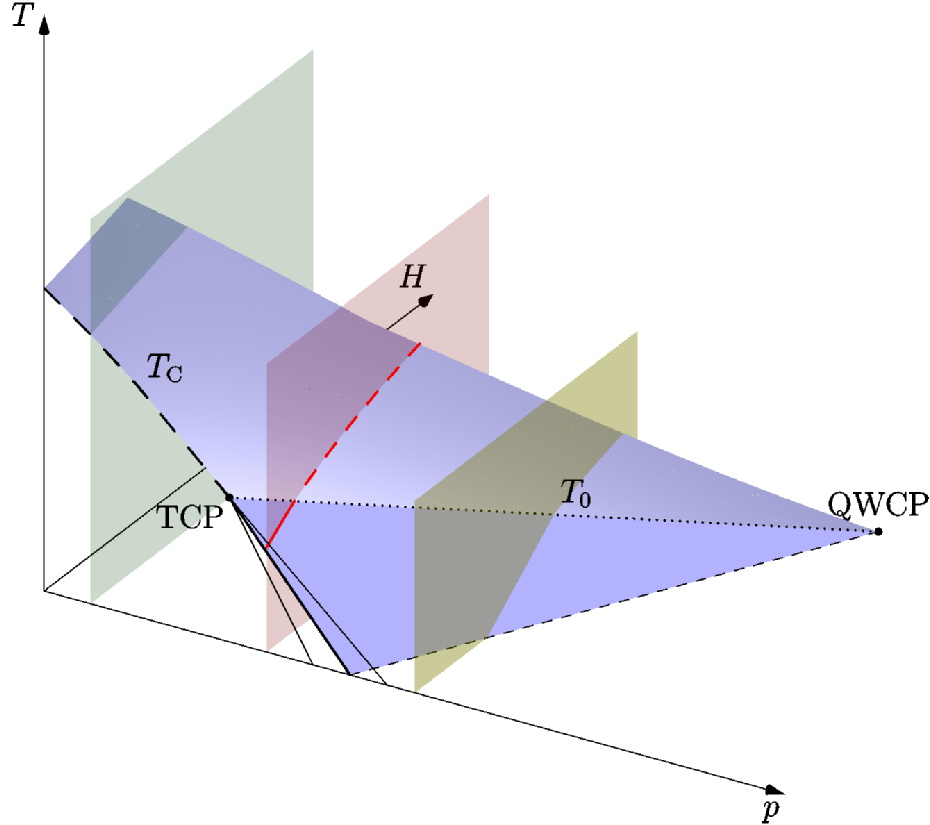}
  \caption{
Schematic view of the proposed phase diagram in the immediate vicinity of the tricritical point (TCP) in the temperature ($T$) -- magnetic field ($H$) -- control parameter ($p$) space based on the experimental data. Within presented study, 
the control parameter is either hydrostatic pressure or substitution. Starting at high temperatures and $H=0$ the second order transition is indicated by dashed 
line, spreading to non-zero field as second order transition plane (light blue surface). Below TCP ($H=0$, solid line) and crossover temperature $T_0$ for $H\neq 0$ (dark blue) the transition becomes first order accompanied by the 
presence of hysteresis (for simplicity indicated only for $H=0$). Planes for fixed $p$ indicate position of samples used in this study --- green (UCo$_{0.990}$Ru$_{0.010}$Al), red (UCo$_{0.995}$Ru$_{0.005}$Al) and brown (UCoAl). 
The red line indicate the phase transition line ($\mu_0 H_c$) for the UCo$_{0.995}$Ru$_{0.005}$Al as shown in Fig.~\ref{F3}, indicating both first-order (solid) and second-order (dashed) transition.
For simplicity the phase boundaries are sketched only for non-negative fields.}
\label{F9}
\end{figure}

The presented experimental data are fairly consistent with the theoretically proposed general phase diagram for itinerant ferromagnets \cite{Belitz2005,Belitz1999} describing the suppression of ferromagnetism as a function of external parameters 
(pressure, composition, see~Fig.~\ref{F9}). Starting from the well-defined ferromagnetic behaviour (1\% Ru doping in this study), phase separation is observed between TCP and QPT both by 0.5\% Ru doping at ambient pressure and 1\% Ru 
doping at hydrostatic pressures between 0.26~GPa and 0.40~GPa. This strongly supports the existence of the expected first-order transition between TCP and QPT.

The presented experimental evidence can be established in several directions. First, we have shown that a very small amount of substitution may not lead to suppression of the first-order transition and consequently the presence 
of a TCP in the phase diagram, but is rather sufficient for tuning the system towards (and across) the QPT. This opens new possibilities for the investigation of QPT in metallic ferromagnets and related phenomena. This would 
suggest a possible way to design and prepare in controlled way a single-crystalline material having the QPT at ambient pressure, unscreened by superconductivity for in-depth studies of related finite-temperature phenomena.

Second, the experimental possibility of smooth passing of the TCP and interchangeability of pressure and composition variations allows us to make a rough estimation of the critical field of the wing-tip QCP.
Following basic thermodynamic considerations within the mean-field model~\cite{Belitz2005}, we obtain a value of about~6~T which is in good agreement with earlier pressure studies on pure UCoAl \cite{Aoki2011} and 
roughly one half of the values reported in recent experimental studies~\cite{Kimura2015}. 

Third, our experimental data shows the importance of the proper and complete investigation of reflections of quantum critical phenomena. The system under investigation (small Ru doping of UCoAl) clearly shows the n=3/2 power law 
in resistivity at low temperatures in paramagnetic ground state, being characteristic for pure transition metal based systems (e.g. archetypal MnSi~\mbox{\cite{Pfleiderer2001}} or ZrZn$_2$~\mbox{\cite{Smith2008}}). This behaviour is not 
fully understood in these systems and probably different mechanisms may be responsible. In the case of small Ru-doping or off-stoichiometric U(Co,Al) one sees the same paramagnetic ground states as in pure UCoAl, however the 
temperature dependence of resistivity may reflect subtle changes in the electron-electron and electron-phonon correlations leading to the striking difference ot the transport properties with respect to those of pure U$TX$ compounds, 
deserving 
detailed investigations both on theoretical and experimental level. Within this context it should be noted, that although the strongly anisotropic U$TX$ (see e.g. discussion on spin fluctuation anisotropy 
in~\mbox{\cite{Mushnikov1999}}) may 
follow~\mbox{\cite{Kirkpatrick2012}} the general features of the magnetic phase diagram~\mbox{\cite{Belitz2005,Belitz1999}} the exact microscopic mechanism behind the non-analytical corrections to the Fermi liquid theory remains 
unknown.

Finally, the possibility of crossing the TCP by means of pressure or composition allows one to perform detailed investigation 
of the basic thermodynamics related to 
phase separation in the immediate proximity of the QPT. This is of particular importance, since contrary to expectations on the basis of basic thermodynamic considerations it is known that phase separation in some quantum 
ferromagnets and other systems can occur away from the coexistence curve of a first-order phase transition~\cite{Kirkpatrick2016}.

\section{Methods}

\subsection{Sample growth and characterization}

Single crystals of UCo$_{1-x}$Ru$_{x}$Al with $x=0.000$, $0.005$, $0.010$ were prepared by the Czochralski method in a tri-arc furnace. High-purity metals were melted into polycrystalline ingot which was turned over 
and remelted to achieve better homogeneity. The single crystals were pulled with a rate of 12 mm/h. The resulting crystals were wrapped in Ta foil and annealed in vacuum (10$^{-6}$~mbar) in a quartz tube for seven days in 
800$^\circ$C. The quality of the single crystals was checked by the X-ray Laue method (Laue diffractometer of Photonic Science), X-ray powder diffraction (XRPD, Bruker AXS D8 Advance X-ray diffractometer with Cu X-rays) and energy-dispersive 
X-ray spectroscopy (EDX, scanning electron microscope Tescan Mira I LMH). Oriented single crystals were cut for specific experiments by a wire saw using solution of silicon carbide powder, glycerin and distilled water. 
Because of the small amount of Ru, inductively coupled plasma optical emission spectroscopy (ICP OES) was used for precise analysis of the resulting composition. Parts of samples with approximate mass of 10~mg were dissolved in HCl and HNO$_3$. 
Optical emission spectroscopy measurements were done using Agilent 5100 ICP-OES equipment. The results of precise analysis by ICP EOS agree very well with the nominal compositions 
(within the 5\% error margin of the method). The residual resistivities of all samples were in the range 10-20~$\mu\Omega$cm allowing us to consider (within the given family of compounds) the obtained samples as candidates for a 
presence of discontinuous QPT~\cite{Brando2016}.

Because of the character of our study, special care was taken to avoid misinterpretation of the obtained data due to the possible inhomogeneity of the samples (multiphase nature, e.g. significant 
variation of the Ru content). This can be fully excluded on the basis of detailed investigations of several pieces cut from different parts of the investigated crystals. As test property, the magnetization, being a very sensitive probe of 
possible variations in composition, was carefully measured as a function of temperature and field.
All tested samples from specific crystals show the same characteristic temperatures ($T_{\rm C}$, $T_0$) and related 
characteristics (remnant magnetization, coercive field, position of the metamagnetic transition in the $H$~--~$T$ plane), justifying the conclusion that the observed behaviour is intrinsic. Moreover, all measurements show single 
well-defined Curie temperatures. This is supported by the result of the pressure experiment on the sample doped with 1\% Ru, which shows stable ferromagnetism with a well-defined Curie temperature and which can be 
driven by hydrostatic pressure to the same 
state as the sample doped with 0.5\% Ru at ambient pressure.

\subsection {Experimental characterization and evaluation}

Magnetic measurements were performed in MPMS 7 XL SQUID magnetometer (Quantum Design) with the magnetic field applied along the c-axis. The magnetization below 1.8~K was measured using Hall probes (Arepoc). For the scaling, the 
measurements from the MPMS magnetometer were used. Measurements under hydrostatic pressure were performed using a CuBe piston cell~\cite{Kamarad2004} suited for  measurement in the MPMS magnetometer. 
The pressure was determined with a Pb manometer, all pressure values are values at low temperatures.
The thermal expansion was measured in a PPMS apparatus ($T>1.8$~K, Quantum Design) and a dilution refrigerator ($T<5$~K, Leiden Cryogenics) using a miniature capacitance dilatometer~\cite{Rotter1998}.

To find the parameters $\rho_0,A,n$ which describes the observed dependence of the resistivity on the temperature, we minimize difference between the observed and the predicted data weighted by the data error in the $L_2$ norm,
i.e. we minimize the following misfit $S$:

$$
S=\frac{1}{2}\left(\mathbf{d}^{\rm pred}-\mathbf{d}^{\rm obs}\right)^{\rm T}\cdot \mathbf{C}_{\mathbf{d}}^{-1}\cdot\left(\mathbf{d}^{\rm obs}-\mathbf{d}^{\rm pred}\right),
$$
where $\bullet^{\rm T}$ is the transposition, $\bullet^{-1}$ is the inversion, $\mathbf{d}^{\rm obs}=\left(R_1,R_2,\dots,R_N\right)^{\rm T}$ is the vector of the observed data, $\mathbf{d}^{\rm 
pred}=\left(\rho_0+AT_1^n,\rho_0+aT_2^n,\dots,\rho_0+AT_N^n\right)^{\rm T}$ is the vector of the predicted data, $N$ is the number of data points. $\mathbf{C}_{\mathbf{d}}$ is the covariance (diagonal) matrix of the data vector,
the diagonal elements are given by $\left(\mathbf{C}_\mathbf{d}\right)_{ii}=\sigma^2_i,\ i=1,\dots,N$, $\sigma_i$ is the error of the $i$-th data point. We estimate that the error of the data points is lower than $1\%$, the
standard error of the $i$-th data point is therefore $\sigma^d_i=0.01R_i$.

In order to obtain the best fit model parameters $\mathbf{m}=\left(m_1,m_2,m_3\right)^{\rm T}=\left(\rho_0,A,n\right)^{\rm T}$ minimizing the misfit $S$, we employ the iterative steepest descent method with the Newton
optimization (see chapter 3.4.2 in~\mbox{\cite{Tarantola2004}}). The uncertainty of the model parameters is determined by the inversion of the Hessian matrix $\mathbf{H}$ (the second-order partial derivatives matrix $H_{ij}=\frac{\partial^2
S}{\partial m_i\partial m_j}$), where model parameters standard errors are computed as follows $\sigma_{\rho_0}=\sqrt{\left(\mathbf{H}^{-1}\right)_{11}}$), $\sigma_{A}=\sqrt{\left(\mathbf{H}^{-1}\right)_{22}}$ and
$\sigma_{n}=\sqrt{\left(\mathbf{H}^{-1}\right)_{33}}$.

\begin{acknowledgments}

This work is a part of the research program GACR 16-06422S which is financed by the Czech Science Foundation. Experiments were performed in the Magnetism and Low Temperature Laboratories, which is supported within the program 
of Czech Research Infrastructures, project no. LM2011025. Authors acknowledge F.R. de Boer for stimulating discussion and careful reading of manuscript.

\end{acknowledgments}


\bibliographystyle{plain}
\bibliography{UCoRuAl}

\end{document}